\documentclass[cits]{PoS}

\title{Large-amplitude intraday variability in QSO 1156+295 observed during a
  VLBI experiment}

\ShortTitle{Large-amplitude intraday variability in QSO 1156+295}

\author{\speaker{Tuomas Savolainen}$^a$ and Yuri Y. Kovalev$^{ab}$\\
        \llap{$^a$}Max-Planck-Institut f\"ur Radioastronomie\\
        Auf dem H\"ugel 69, 53121 Bonn, Germany\\
        \llap{$^b$}Astro Space Center of Lebedev Physical Institute\\
        Profsoyuznaya 84/32, 117997 Moscow, Russia\\
        E-mail: \email{tsavolainen@mpifr-bonn.mpg.de},
                \email{ykovalev@mpifr-bonn.mpg.de}}

\abstract{

  We present here the discovery of rapid, large amplitude intraday variability
  in the compact flat-spectrum radio quasar 1156+295. The detection of 40\%
  flux density variations at 15\,GHz on a timescale of only 2.7\,hours was
  serendipitously made when the source was observed with the Very Long
  Baseline Array as a part of the MOJAVE survey programme on February 5,
  2007. Intraday variability on timescales of a few hours or less is rare, and
  there exist very few sources that show large-amplitude variations on a
  timescale as short as what is now observed for 1156+295. The shape of the
  visibility function of the source changes very little during the
  observation, although the correlated flux density changes by 40\%. This
  suggests that the variability occurs in a single dominant compact
  component. The observed variability characteristics are consistent with
  interstellar scintillation in nearby, highly turbulent medium. The rms
  amplitude of modulation at 15\,GHz is unusually large and it implies a
  rather high scattering measure along the line-of-sight towards 1156+295.

}

\FullConference{Workshop on Blazar Variability across the Electromagnetic Spectrum\\
		 April 22-25 2008\\
		 Palaiseau, France}

\begin{document}

\section{Introduction}

Flux density variations on timescales of $\lesssim2$ days -- so-called
intraday variability (IDV) -- are seen at centimetre wavelengths in a
significant fraction of compact, flat-spectrum AGN \cite{wag95,lov03}. There
is now good evidence that, at least in the case of the most extreme sources
showing large-amplitude variations on intra-hour timescales, the IDV is due to
interstellar scintillation (ISS) in the turbulent, ionised interstellar
medium\footnote{We note that there are some sources likely exhibiting also
  intrinsic variability on a timescale of $\sim1$\,day \cite{qui91}.}  (see
e.g. \cite{big07} and references therein). This evidence is mainly based on
the detection of time delays in the variability pattern arrival times between
widely separated telescopes and on the observations of an annual modulation of
the variability timescale. By detailed analysis of the scintillation it is
possible to probe both the ISM and the structure of compact radio sources at
microarcsecond resolution, which is far higher than what can be achieved by
the present day VLBI \cite{mac02}. The short variability timescale of the most
extreme sources facilitates such studies by allowing well-sampled observations
of the flux density fluctuations to be made within a single observing
run. Unfortunately, however, while IDV in flat-spectrum radio sources is
common, variability on timescales of a few hours or less and with an rms
amplitude of modulation larger than 10\% is extremely rare and only a handful
of such sources are known. We report here the serendipitous discovery of one
new source, QSO 1156+295, showing IDV on a timescale of less than 3\,h and
with 13\% rms amplitude of modulation. This discovery is unusual in two
ways. Firstly, it was made during a VLBI experiment, and secondly, the
modulation index of 13\% is atypically high at 15\,GHz.

\begin{figure}
\centering
\includegraphics[angle=-90,width=0.45\textwidth]{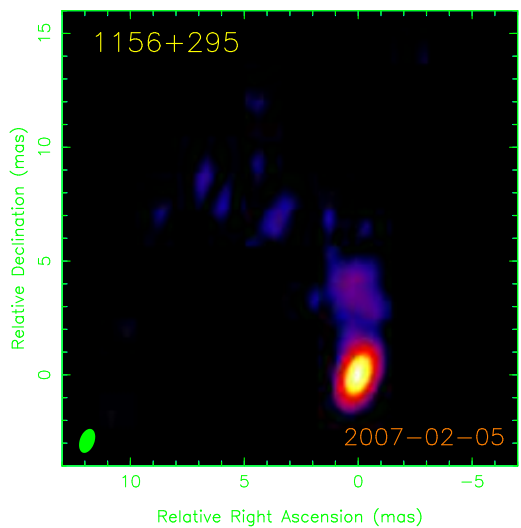}
\caption{Naturally weighted 15\,GHz false colour image of 1156+295 observed
  with the VLBA on February 5, 2007. The map peak brightness is 1.46 Jy
  beam$^{-1}$ and the low flux density cut-off is at 0.7 mJy beam$^{-1}$.}
\label{map}
\end{figure}

\section{Observations and analysis}

1156+295 (4C +29.45) is a quasar at $z=0.729$ that is strongly variable
throughout its spectrum from radio to gamma-rays \cite{kov02,wil83,sre96}. It
has a core-jet morphology in VLBI maps and it appeared strongly core-dominated
in 2007 (Fig.~\ref{map}). 1156+295 was observed with the NRAO's Very Long
Baseline Array (VLBA) at 15\,GHz as a part of the MOJAVE project \cite{lis05} on
February 5, 2007. In the same 24-hour session, 24 other sources beside
1156+295 were also observed. The individual scans of different sources were
interleaved in order to maximise the $(u,v)$ coverage for each
source. 1156+295 was observed for 9 scans distributed between $4^h$ and
$16^h$\,UT and each lasting 4.7\,min. The data were calibrated using the
standard methods of VLBI data reduction (see description in
\cite{lis05}). After \emph{a priori} amplitude calibration and fringe fitting,
it was noticed that the correlated flux density of 1156+295 shows strong,
correlated temporal variability from scan to scan at \emph{every
  baseline}. There is a deep minimum, surrounded symmetrically by two maxima,
in the correlated flux density curves between $7^h$ and $12^h$\,UT. The
peak-to-trough amplitude of this dip is 0.6\,Jy, 40\% of the average flux
density, and the variability timescale is only 2.7\,h. The dip can be seen at
every baseline, which excludes the source structure as the cause of the
variability. Also, neither the system temperature measurements used in the
amplitude calibration nor the fringe-fit solutions of the experiment showed
any anomalies that could explain the variability as an instrumental effect.

\begin{figure}
\includegraphics[width=\textwidth]{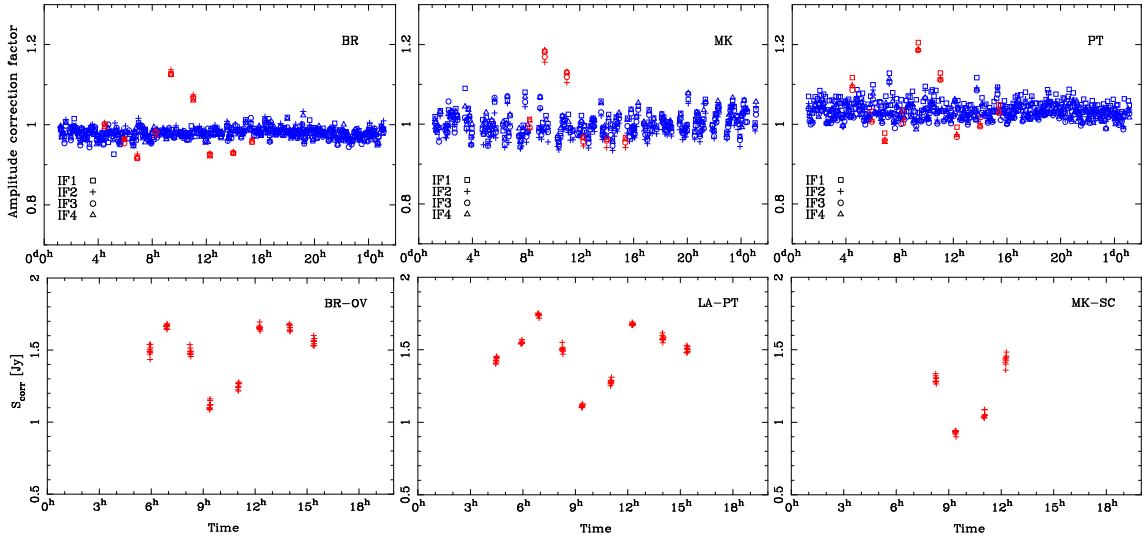}
\caption{\emph{Top:} Amplitude self-calibration solutions of three example
  antennas (BR, MK, and PT) for the VLBA observation on February 5, 2007. The
  figure shows one solution per scan and per IF. Scans at elevations below
  $15^\circ$ are excluded. Solutions for 1156+295 are shown in red colour,
  while blue colour corresponds to the solutions of all the other sources. The
  obvious discrepancy between the solutions for 1156+295 and for the other
  sources is present at all ten antennas. \emph{Bottom:} Examples of the
  correlated flux density curves of 1156+295 at three baselines after
  calibration of antenna gains by amplitude correction factors interpolated
  from the nearby scans of other sources. Interpolation was done from the
  scans that were within 1\,h of the target scan, and that were observed at
  source elevation $>15^\circ$. Time is in UT.}
\label{baseline}
\end{figure}

We tested the reality of the flux density variations by imaging and
self-calibrating the $(u,v)$ data of 1156+295 in a usual manner and comparing
the resulting antenna gain correction factors with those derived from the
self-calibration of the other 24 sources observed in the same experiment. The
top panels in Fig.~\ref{baseline} show these gain amplitude correction factors
for three example antennas. While the amplitude self-calibration with a
5-minute solution interval is able to remove the variability in 1156+295, it
results in gain amplitude correction factors that are -- \emph{for every
  antenna} -- significantly offset from those determined from the 24 other
sources. This can be explained if the observed variability is indeed genuine
IDV, but does not significantly change the shape of the source's visibility
function during the observation. This happens for example when the variability
occurs in a single dominant, compact component containing most of the source's
flux. Calibration errors cannot account for the behaviour described above,
because the antenna gain is source-independent. Thus, we conclude that the
variations in the correlated flux density of 1156+295 are due to genuine IDV.

\begin{figure}
\includegraphics[width=0.495\textwidth]{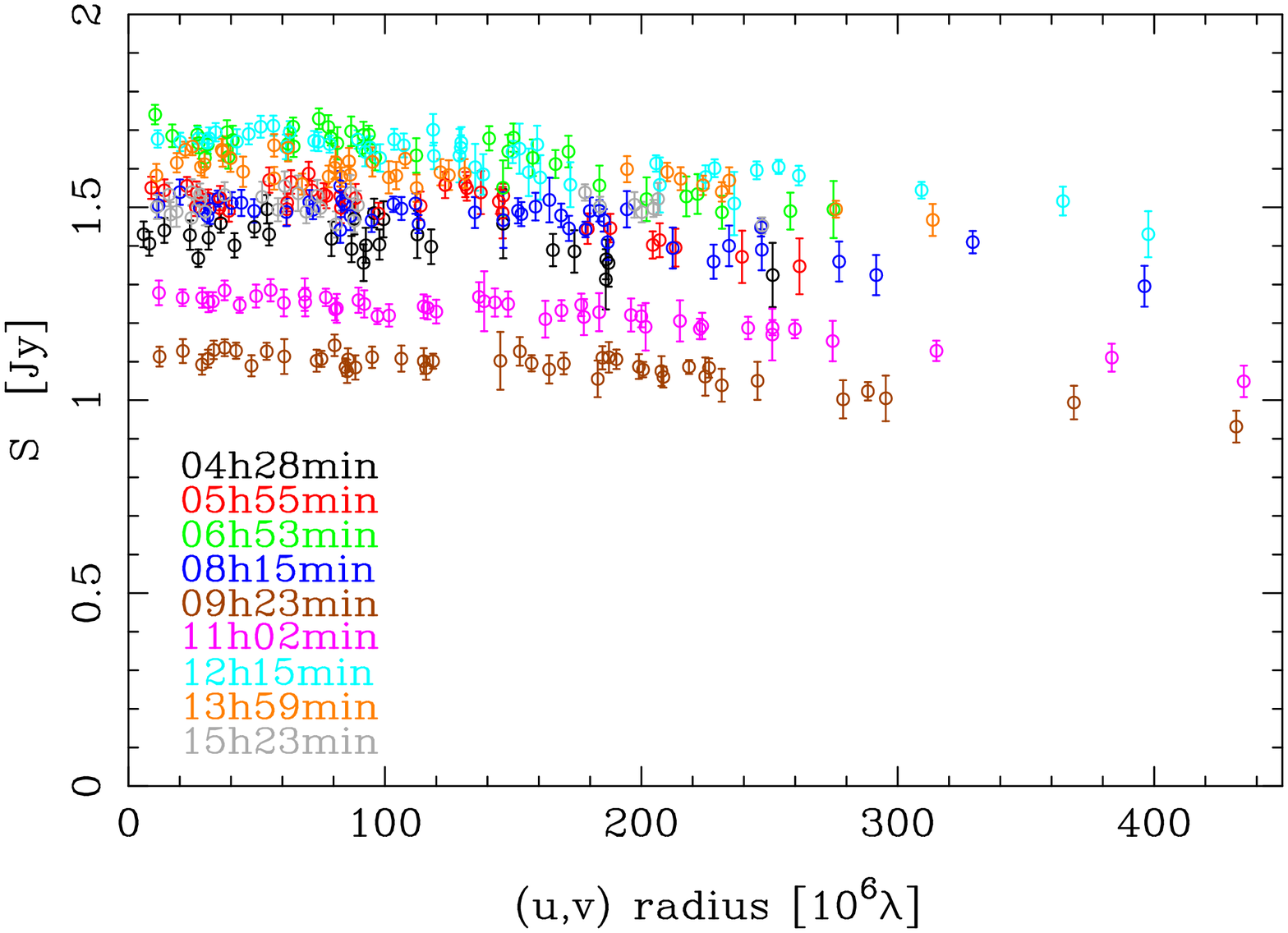}
\includegraphics[width=0.495\textwidth]{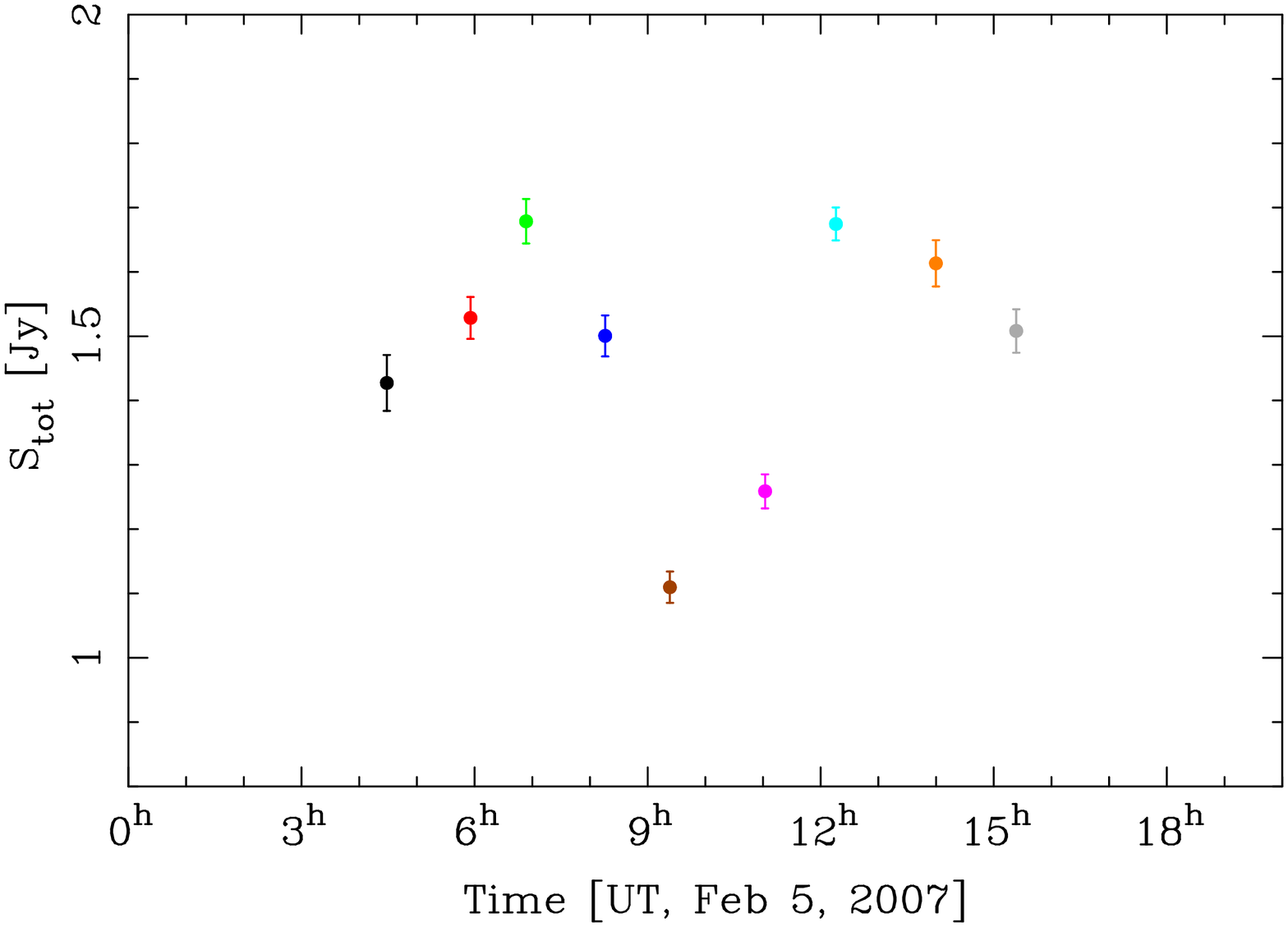}
\caption{\emph{Left:} Calibrated correlated Stokes I flux density of 1156+295
  as a function of $(u,v)$ radius for each individual scan of the VLBI
  experiment on February 5, 2007. The data has been averaged over the IFs and
  over the scan length in time. The different scans are shown in different
  colours with the scan start time in UT indicated in the legend.  
  \emph{Right:} Integrated flux density curve of 1156+295, obtained by
  averaging the visibility amplitudes at projected baselines shorter than
  100\,M$\lambda$. The colour-coding of the scans corresponds to the one used
  in the left panel.}
\label{lightcurve}
\end{figure}

\begin{figure}
\centering
\includegraphics[width=0.495\textwidth]{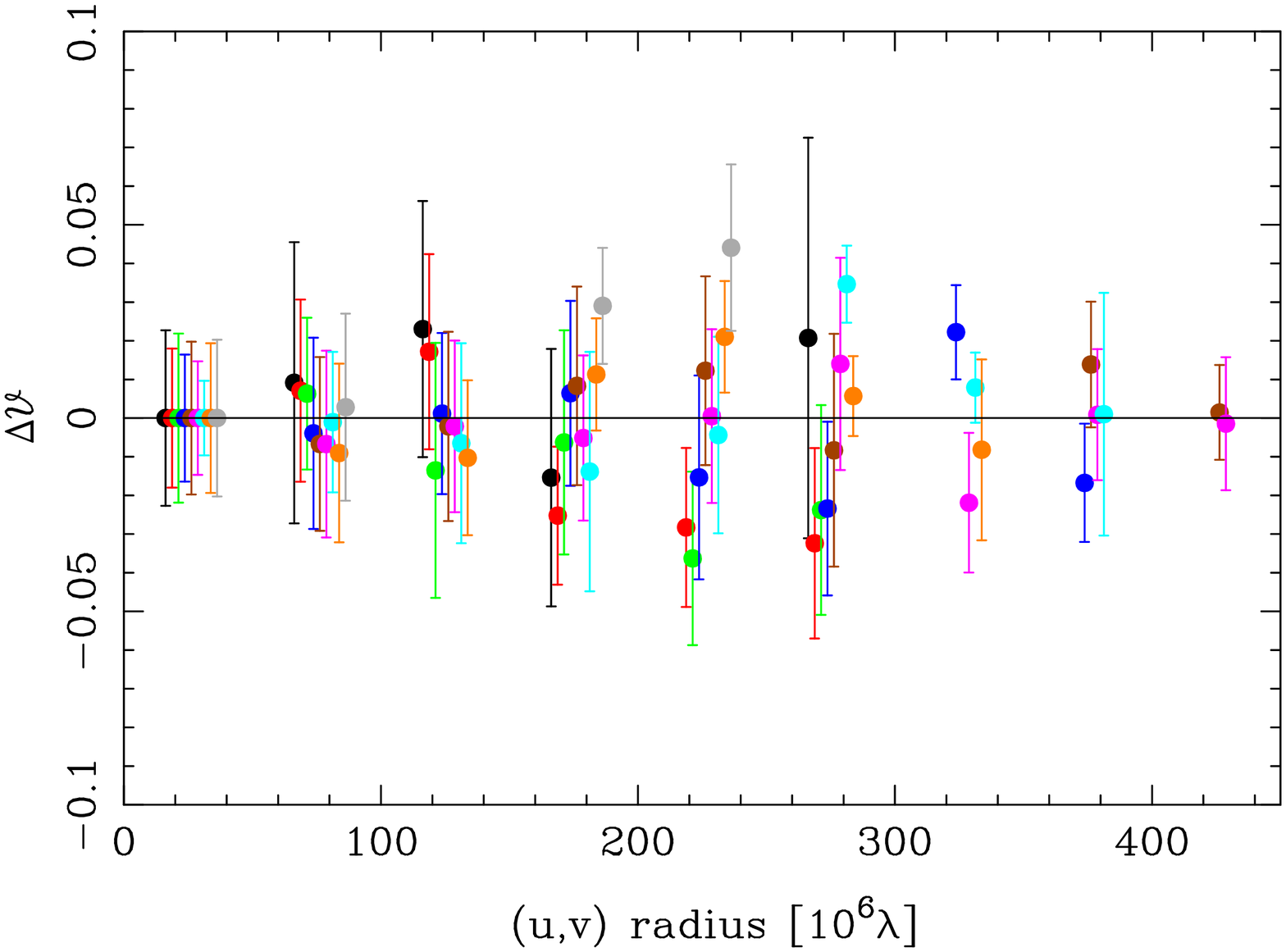}
\includegraphics[width=0.495\textwidth]{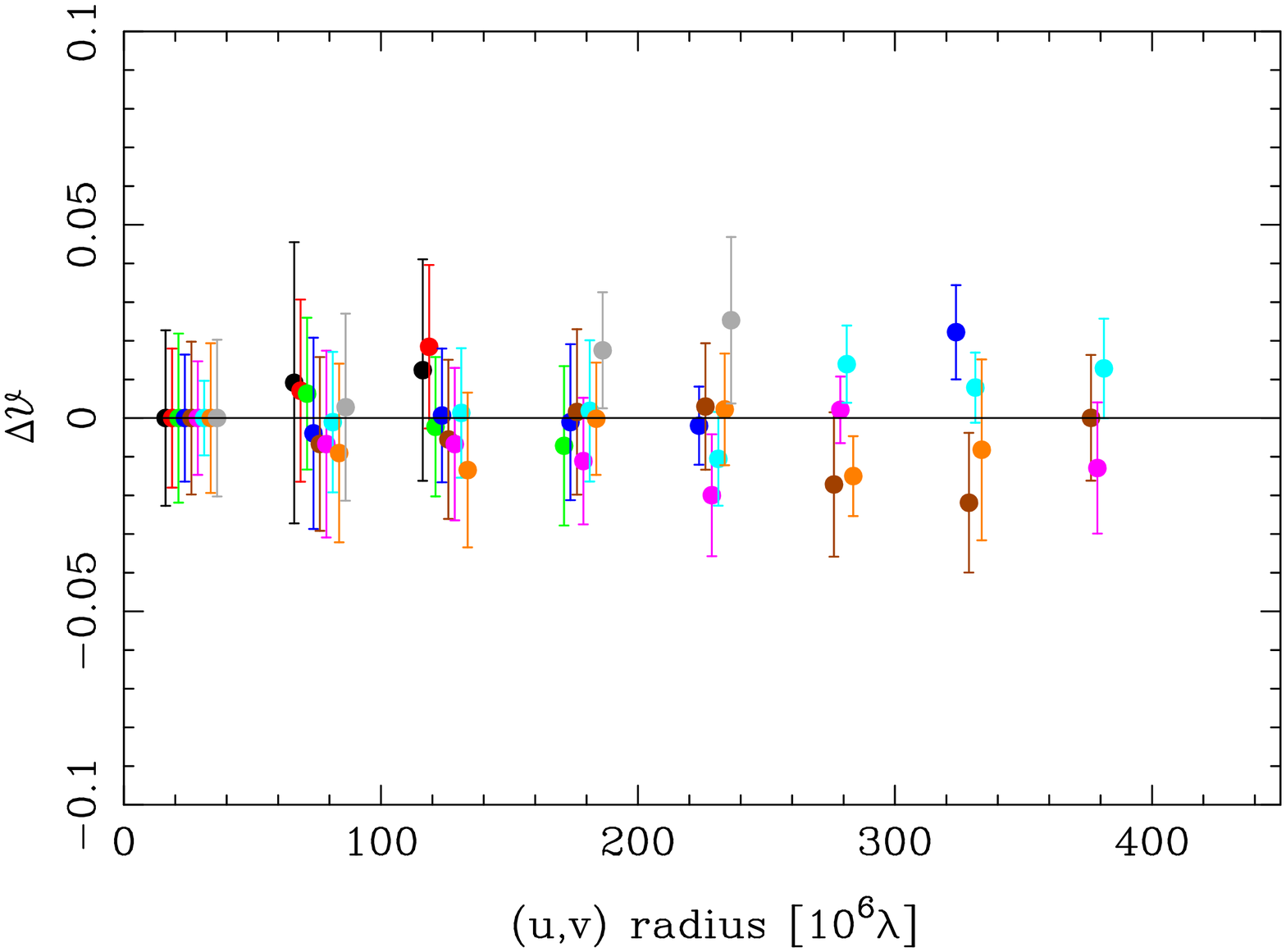}
\caption{\emph{Left:} Difference between the visibility function of each
  individual scan of 1156+295 and the average, binned visibility function of
  the whole observation. The visibility function is obtained by normalising
  the binned correlated flux density of each scan to 1.0 in the first bin. The
  bin size is 50\,M$\lambda$. The scans are shown with the same colour-coding
  as in Fig.~3. As can be seen, there is little change in the normalised
  visibility function despite the strong variability of the correlated flux
  density. During the observation, the normalised visibility function changes
  at $(u,v)$ radius of $200-300$\,M$\lambda$ by about $8$\%. \emph{Right:}
  Same as in the left panel, but without baselines to St. Croix. Now the small
  changes in the shape of the visibility function during the observation have
  completely vanished.}
\label{norm}
\end{figure}

The residual errors in the antenna gains, which remain after the \emph{a
  priori} amplitude calibration, need to be corrected before the observed
variability in 1156+295 can be properly analysed. The large number of sources
that were observed during the MOJAVE session allow us to estimate the gain
amplitude correction factors for 1156+295 by interpolating the
self-calibration solutions from the nearby scans of other sources. Examples of
correlated flux density curves that were calibrated in this way are shown in
the bottom panels of Fig.~\ref{baseline}. As can be seen in the figure, the
variations in the correlated flux density are very similar at long (MK-SC),
intermediate (BR-OV), and short (LA-PT) baselines. The left panel of
Fig.~\ref{lightcurve} shows the calibrated correlated flux density of 1156+295
as a function of $(u,v)$ spacing for each individual scan. The figure suggests
the shape of the visibility function does not change appreciably during the
observation, although the flux density changes by 40\%. To analyse possible
small changes, we have plotted in Fig.~\ref{norm} the difference between the
normalised visibility function of each scan and the average visibility
function of the whole observation of 1156+295. There are changes, on the level
of $<8$\%, in the shape of the visibility function during the observation.
However, as can be seen in the right panel of Fig.~\ref{norm}, these changes
are purely due to the baselines to one antenna (St. Croix), and they can be
caused by calibration inaccuracies of this antenna. We are therefore unable to
conclusively confirm the detection of time variability in the shape of the
visibility function. An upper limit to variability is 8\% during the course of
observation.

The right panel of Fig.~\ref{lightcurve} shows the integrated VLBA flux
density curve for 1156+295, which we constructed by averaging the visibility
amplitudes at projected baselines between $6-100$\,M$\lambda$. Since the
visibility function is essentially flat in this range (see the left panel of
Fig.~\ref{lightcurve}), the averaging gives a good estimate of the integrated
emission coming from angular scales $\lesssim30$\,mas. Averaging also reduces
the errors that are due to inaccurate amplitude calibration, since these are
antenna-specific. From the integrated VLBA flux density curve, we calculate
the variability timescale as the average of the peak-to-trough and
trough-to-peak times of the big dip, which gives
$t_\mathrm{var}=2.7\pm0.5$\,h. The modulation index $m$, defined as the
standard deviation of the flux density curve divided by the mean flux density,
is $13\pm3$\%. This value of $m$ is much higher than what is typically
observed for IDV sources at frequencies above 5\,GHz \cite{ked01,kra03}.

\section{Discussion}

Through light-travel time arguments, the variability timescale of 2.7\,hours
translates into a brightness temperature of $\gtrsim 2 \times 10^{19}$\,K for
1156+295 if the observed flux density variations are source-intrinsic. This is
far in excess of the inverse Compton (IC) catastrophe limit of $10^{12}$\,K
\cite{kel69}. Therefore, a very high Doppler factor would be needed in order
to explain the observations in the standard framework of incoherent
synchrotron radiation from a jet of relativistic electrons. A Doppler factor
of $\gtrsim 270$ would be needed in order to avoid the IC catastrophe in the
case of a power-law electron energy distribution, and even in the case of the
quasi-monoenergetic electron population \cite{kir06}, a Doppler factor of
$\gtrsim200$ would be needed. Such fast jets are unlikely to exist on both
theoretical \cite{beg94} and observational \cite{coh07} grounds. Thus, we
consider it probable that the fast variability observed in 1156+295 is due to
source-extrinsic effects.

A sufficiently compact radio source will scintillate due to the wave
propagation effects in the ionised interstellar medium of our Galaxy (for a
review, see e.g. \cite{ric90,goo97}). The scattering strength is determined by
the strength of the electron density fluctuations along the line of
sight. Below a certain critical frequency $\nu_\mathrm{s}$, both narrow-band
diffractive and broad-band refractive scattering phenomena can be observed and
the scattering is referred to as ``strong''. Above the critical frequency, in
the so-called ``weak'' scattering regime, the refractive and diffractive
scattering length scales become equal and flux density variations arise due to
slight focusing and defocusing over the Fresnel scale.

The variability timescale and modulation index can be used to constrain the
properties of the scattering medium. IDV has been previously observed in
1156+295 at 5\,GHz in 2002, albeit with less extreme characteristics:
$m=5.8$\% and $t_\mathrm{var}\sim20$\,h have been reported \cite{lov03}. For a
point source, $m \propto \nu^{17/30}$ and $t_\mathrm{var} \propto \nu^{-11/5}$
in the case of strong refractive ISS, while in the weak regime $m \propto
\nu^{-17/12}$ and $t_\mathrm{var} \propto \nu^{-1/2}$ \cite{ric90}. Therefore,
the difference between timescales observed at 5\,GHz and 15\,GHz suggests that
$\nu_\mathrm{s}$ is above 5\,GHz. This implies a rather large scattering
measure towards 1156+295, which is somewhat unexpected considering the high
galactic latitude ($b=78.4^\circ$) of the source \cite{cor02}. However, if the
scintillation occurs in the very local ISM, there is not necessary a strong
correlation between the scattering measure and the galactic latitude
\cite{bha98}. There are also 5 years between these observations, and the
source could have been in a more compact stage at our epoch than in
2002. Simultaneous multifrequency measurements are needed to settle this.

Setting a lower limit to the intrinsic source size,
$\theta_\mathrm{s}^{FWHM}$, by the IC catastrophe argument, it is possible to
constrain the maximum distance to the scattering screen. If the maximum
Doppler factor $\delta$ is assumed to be $<50$ \cite{coh07}, the lower limit
to $\theta_\mathrm{s}^{FWHM}$ is about 17\,$\mu$as. Together with the short
variability timescale of 2.7\,h and the modulation index of $13$\%, this
implies a maximum screen distance of about 300\,pc for typical screen
velocities \cite{goo97,sav08}. If equipartition conditions in the source are
assumed, we can estimate that $\theta_\mathrm{s}^{FWHM} \sim 270 \cdot
\delta^{-7/17}$\,$\mu$as. Again assuming $\delta < 50$, this would place the
screen at the distance of $\lesssim 100$\,pc. The above calculations constrain
also the scattering measure \cite{goo97}, which indeed turns out to be rather
large: {\it SM} $\gtrsim 0.5 $\,m$^{-20/3}$\,pc for a screen distance of
$\lesssim 300$\,pc, and {\it SM} $\gtrsim 4$\,m$^{-20/3}$\,pc for a screen
distance of $\lesssim 100$\,pc (an uncertainty of 3\% in $m$ has been assumed
when calculating these limits; a more detailed analysis is presented in
\cite{sav08}).

The rapid, large amplitude IDV in 1156+295 is in principle consistent with
interstellar scintillation due to a nearby, localised region of highly
turbulent ionised gas. The above-derived lower limits for {\it SM} in the
direction of 1156+295 are, however, 5-40 times larger than what is predicted
by Cordes \& Lazio \cite{cor02} model of the distribution of free electrons in
the Milky Way. Therefore, further observations should be carried out to search
for other signs of increased turbulence in the direction of 1156+295. Finally,
we note that the case of 1156+295 is also an important reminder that large
amplitude IDV within the timescale of a VLBI experiment violates the basic
assumption made in Earth rotation synthesis, and can significantly degrade
VLBI imaging results.

\acknowledgments 

We thank Lars Fuhrmann, David Jauncey, Thomas Krichbaum, Matthew Lister and
Sarma Kuchibhotla for useful discussions. Both authors are research fellows of
the Alexander von Humboldt Foundation. TS was also partially supported by the
Max-Planck-Gesellschaft and by the Academy of Finland grant 120516. The
National Radio Astronomy Observatory is a facility of the National Science
Foundation operated under cooperative agreement by Associated Universities,
Inc.

\end{document}